\documentclass[a4paper]{JHEP3}
\usepackage{amsmath}
\usepackage{amsthm}
\usepackage{amsfonts}          % if you want the fonts
\usepackage{amssymb}           % if you want extra symbols
\usepackage[all]{xy}           % Commutative diagrams
\usepackage{stmaryrd}          % \lightning is defined there
\usepackage{afterpage}

\usepackage{graphicx}

\newcommand{\diff}{{\rm d}}
\newcommand{\Span}{\mathop{\rm span}\nolimits}
\newcommand{\eqdef}{\stackrel{\rm def}{=}}

\newcommand{\C}[1]{\ensuremath{{\mathbb{C}^{#1}}}}

\newcommand{\Z}[1]{\mathbb{Z}_{#1}}

\newcommand{\re}{\mathop{\rm Re}\nolimits}
\newcommand{\im}{\mathop{\rm Im}\nolimits}

\newcommand{\tr}{\mathop{\rm tr}\nolimits}

\newcommand{\End}{{\mathop{\rm End}\nolimits}}

\newcommand{\Dbrane}{D--brane}
\newcommand{\Dbranes}{D--branes}
\newcommand{\CY}{Calabi--Yau}
\newcommand{\CYm}{\CY{} manifold}
\newcommand{\CYms}{\CYm{}s}

\newcommand{\Ktheory}{K--theory}

\newcommand{\Kgroup}{K--group}
\newcommand{\Kgroups}{K--groups}

\newcommand{\Ncal}{\mathcal{N}}
\newcommand{\Db}{{\mathop{{}\rm D}\nolimits}^b}
\newcommand{\Coh}{{\mathop{\rm Coh}\nolimits}}

\newcommand{\rep}{{\mathop{\rm rep}\nolimits}}
\newcommand{\module}{{\mathop{\rm mod}\nolimits}}

\newcommand{\catrep}{\hbox{-}\rep}
\newcommand{\catmod}{\hbox{-}\module}

\newcommand{\T}[1]{\lbrack{#1}\rbrack} % shift functor X[i]
\newcommand{\Hom}{{\mathop{\rm Hom}\nolimits}}
\def\gldim{\qopname\relax{no}{gldim}}

\newcommand{\qnode}[1]{\hbox{% put everything into a hbox
\rlap{% step1: draw circle with zero width & height 
\kern -0.3mm% small adjustment
\lower 1.35mm \hbox{\Huge $\circ$}}% Draw fat circle
\rlap{% step2: enter text with zero width & height
\hbox to 3.9mm{\hfill$\scriptstyle #1$\hfill}}% text
\phantom{$\bigcirc$}% step3: get space from invisible bigcirc
}% end of hbox
}% end of newcommand

\newcommand{\textdef}[1]{\textbf{#1}}

\newsavebox{\QuivToricA}
\newsavebox{\QuivToricB}
\newsavebox{\QuivToricC}

\newenvironment{descriptionlist}{%
\begin{list}%
{}%
{}}%
{\end{list}}

\newenvironment{displayarray}%
{\everymath{\displaystyle\everymath{}}\array}%
{\endarray}

\newtheorem{theorem}{Theorem}
\newtheorem{lemma}{Lemma}
\newtheorem{example}{Example}

\newtheorem{conjecture}{Conjecture}

\newtheorem{definition}{Definition}

\title{On Berenstein--Douglas--Seiberg Duality}

\author{
  Volker Braun\\
  Ecole Normale Sup\'erieure, 24 rue Lhomond, 75231 Paris, France\\
  E-mail: \email{volker.braun@lpt.ens.fr}
}

\preprint{
  hep-th/0211173 \\
  LPTENS-02/58
}

\abstract{I review the proposal of Berenstein--Douglas for a
  completely general definition of Seiberg duality. To give evidence
  for their conjecture I present the first example of a physical dual
  pair and explicitly check that it satisfies the requirements. Then I
  explicitly show that a pair of toric dual quivers is also dual
  according to their proposal. All these computations go beyond
  tilting modules, and really work in the derived category. I
  introduce all necessary mathematics where needed.  }


\begin{document}

\section{Introduction}

There exists a wealth of knowledge about $\Ncal=1$ field theories (see
e.g.~\cite{BermanRabinovici}), none of which can be proven in any
rigorous sense. Of course the handwaving nature of the arguments makes
it hard to verify or falsify anything. Yet recently Berenstein and
Douglas~\cite{BerensteinDouglas} proposed an exact criterion to decide
whether two theories can be (generalized) Seiberg dual. Their proposal
is elegant to formulate yet not so obvious to check in practice. The
purpose of this paper is to study the implications for physical
quivers, as opposed to the toy model in~\cite{BerensteinDouglas}.

Since there is no useful introduction I review their proposal and
explain the ideas involved. I also revisit the example that they give
to provide evidence for their conjecture.

While computationally simple, their example suffers from a serious
illness: It requires F-term constraints which do not come from a
superpotential, so if at all it can only be a subsector of the full
quiver. To remedy this I exhibit a sample pair of quivers which are
completely physical.

I then set out to explicitly check that my example is (generalized)
Seiberg dual according to the proposal. It turns out that one really
has to work in the derived category, i.e. use tilting complexes
instead of just tilting modules. Because this is certainly not
standard knowledge amongst physicist I carefully explain how to do
this in an attempt to make everything self-contained (for another very
nice introduction see~\cite{HolmSummerSchool}).

After a lightning review of toric duality I then explicitly check that
one pair of toric dual quivers is BDS-dual. This works in the same way
as in the previous example, but is technically more challenging.

Finally I discuss some necessary conditions for two quivers to be dual
which can sometimes be understood from physical intuition. I prove
that one can not simply ``fix'' the example
of~\cite{BerensteinDouglas} by adding an additional arrow.

\section{Review of the BDS--Duality Conjecture}

\subsection{Beyond Representations}

The duality conjecture is based on properly distinguishing between the
``theory'' and its representation. Of course physicists in general
ignore such subtleties, not completely without reason (e.g. group
theory vs. representation theory of groups). However in this case it
is important to make this distinction.

As is well-known, the $\Ncal=1$ SYM theory for $n$ gauge groups and
$k$ bifundamental chiral multiplets is defined by the Lagrangian
\begin{equation}
  \label{eq:N1SYM}
  \begin{split}    
  \mathcal{L} =& \im \tr \left[
    \sum_{i=1}^n
    \int \diff^2 \theta \,
     W_{\alpha}^{(i)} W^{(i), \alpha}
  + 
    \sum_{j=1}^k
    \int \diff^2 \theta \diff^2 \bar{\theta} \,
     \Phi^{(j), \dag}
     e^{V^{(j\text{-fund})}}
     \Phi^{(j)}
     e^{-V^{(j\text{-anti-fund})}}
  \right] +
  \\ 
  &+ 
%    \sum_{i=1}^n 
%    \int \diff^2 \theta \diff^2 \bar{\theta} \,
%    \zeta_i V^{(i)}
%  +
    \re \int \diff^2 \theta \,
    W(\Phi_1,\dots,\Phi_k)
  \end{split}
\end{equation}
Given the Lagrangian one can expand it in component fields, write down
F- and D-term constraints, derive Feynman rules etc. However there is
one piece of information missing if one wants to compute actual
numbers instead of algebraic manipulations: Nothing
in~\eqref{eq:N1SYM} tells you the ranks of the gauge groups\footnote{I
  will restrict myself to unitary gauge groups throughout this paper}
$U(N_i)$ and the vevs of the bifundamental fields. This choice of
dimensions for vector spaces and explicit matrices is the
representation data that we need to completely specify the physics.

This distinction is exactly mirrored by the theory of quivers and
their representations. Recall that you can write~\eqref{eq:N1SYM} in
the following graphical notation:
\begin{enumerate}
\item For each factor of the gauge group draw a node.
\item For each bifundamental chiral superfield draw an arrow. The
  field transforms in the antifundamental of the $i$th factor of the
  gauge group and the fundamental of the $j$th, and the arrow is
  directed from $i$ to $j$.
\item For each F-term constraint $\frac{\partial W}{\partial \phi_i}$
  define a relation for the corresponding arrows.
\end{enumerate}
The resulting directed graph with relations (quiver) encodes all the
information in~\eqref{eq:N1SYM}.

A quiver representation is a choice of vector space for each node, and
a choice of matrix for each arrow such that the product of the
matrices satisfies the given relations. In other words it is precisely
the representation data that has to complement the
Lagrangian~\eqref{eq:N1SYM}. 

The abstraction from representation theory data in SYM theories and
quivers is illustrated in table~\ref{tab:theoryrepresentation}.
\begin{table}[htbp]
  \centering
  \begin{tabular}{c||c|c}
    & Algebraic data 
    & Representation data 
    \\ \hline\hline
      \txt{$\Ncal=1$\\SYM} 
    & 
      \begin{minipage}[c]{55mm}
      \begin{center}
        \medskip
        \begin{math} 
        \begin{array}{c}\displaystyle
          \mathcal{L}= 
          \im \tr \sum_{i=1}^3
          \int \diff^2 \theta \,
%         W_{\alpha}^{(i)} W^{(i), \alpha}
          \big(W^{(i)}\big)^2
          +
        \hfill \\ \displaystyle \hfill
          + \cdots + 
          \re \int \diff^2 \theta  \, W
        \end{array}
        \end{math}
        \\
        $W=\tr(\phi_{12}\phi_{23}\phi_{31})$
        \medskip
      \end{center}
      \end{minipage}
    & 
      \begin{minipage}[c]{55mm}
      \begin{center}
        Gauge group~$U(1) \times U(2) \times U(1)$ \\
        \begin{math}
        \begin{array}{r@{=}l}
          \left<\phi_{12}\right> &
          \left(
            \begin{smallmatrix} 1 & 0 \end{smallmatrix}
          \right) \\
          \left<\phi_{23}\right> &
          \left(
            \begin{smallmatrix} 0 \\ 1 \end{smallmatrix}
          \right) \\
          \left<\phi_{31}\right> &
          \left(
            \begin{smallmatrix} 0 \end{smallmatrix}
          \right)
        \end{array}
        \end{math}
      \end{center}
      \end{minipage}
    \\  \hline
      Quiver 
    & 
      \begin{minipage}[c]{55mm}
      \begin{center}
        \begin{math}
        \vcenter{\xymatrix@R=0.8cm@C=0.3464cm{ & 
          \qnode{3} \ar[ld]_{\phi_{31}} \\
          \qnode{1} \ar[rr]^{\phi_{12}} & & 
          \qnode{2} \ar[lu]_{\phi_{23}} 
        }}
        \begin{array}{c}
          \phi_{12} \phi_{23} =0 \\ 
          \phi_{23} \phi_{31} =0 \\ 
          \phi_{31} \phi_{12} =0 
        \end{array}
        \end{math}      
      \end{center}
      \end{minipage}
    & 
      \begin{minipage}[c]{55mm}
      \begin{center}
        \begin{math}
        \vcenter{\xymatrix@R=0.8cm@C=0.3464cm{ & 
          \C{}  \ar[ld]_{\left(
                           \begin{smallmatrix} 0 \end{smallmatrix}
                         \right)} \\
          \C{}  \ar[rr]^{\left(
                           \begin{smallmatrix} 1 & 0 \end{smallmatrix}
                         \right)} & & 
          \C{2} \ar[lu]_{\left(
                           \begin{smallmatrix} 0 \\ 1 \end{smallmatrix}
                         \right)} 
        }}      
        \end{math}      
      \end{center}
      \end{minipage}
    \\
  \end{tabular}
  \caption{Examples for algebraic vs. representation data}
  \label{tab:theoryrepresentation}
\end{table}
Two remarks are in order: First note that the relations of the quiver
stand on the same footing as the directed graph. Drawing the graph and
not writing down the relations is as nonsensical as leaving out random
arrows or nodes. Second, I have ignored the D-term constraints. They
are of course important, and are reflected by the D-flatness
conditions on the SYM side and the choice of ``stability'' on the
quiver side. However they will play no role in the following, so I
will not review this here.

Now fix one SYM Lagrangian resp. quiver, then it is obviously
interesting to understand the set of possible representations. The
all-important observation for the following is that there is an
obvious notion of ``map'' from one representation to the other. Such a
map is linear on the vector space attached to each node, and each
square induced by an arrow has to commute. This is the straightforward
generalization of the idea that one representation is a
subrepresentation of another if it is contained as a block in the
defining matrices. This notion of ``map'' is well-known in the
representation theory of quivers, and turns the set\footnote{As usual
  we assume that all categories are small.} of representations into
the category of representations. All that seems rather trivial, but
being able to do ``linear algebra'' ultimately leads us to the derived
category of such representations.

\subsection{The Conjecture}

We want to understand Seiberg duality, that is in the simplest case
$U(N_c)$ dual to $U(N_f-N_c)$ + superpotential. So it connects two
representations of different theories. 

The new approach of~\cite{BerensteinDouglas} is to ask what remains on
the purely algebraic level, without the representation theory data.
Surely the dual Lagrangians must be somehow connected, although their
representation theory is very different, one is defined for all $N_f$,
$N_c$ and the other only for $N_f>N_c$.

In the language of quivers, a Seiberg dual pair $Q_1$, $Q_2$ has
different representation categories: $Q_1\catrep \not= Q_2\catrep$. It
is not at all obvious what $Q_1$ and $Q_2$ should have in common.

We need some additional input. Consider Type IIB string theory
compactified on some \CY{} $3$-fold, and add \Dbranes{} preserving
half of the supersymmetry. Then one gets a $\Ncal=1$ low energy
effective field theory, and we can try to study the field theories by
analyzing the different \Dbrane{} configurations.

Now it was argued that the correct framework for studying the
\Dbrane{} configurations is $\Db(\Coh X)$, the bounded derived
category of coherent sheaves on the \CYm{} $X$. This would be a
terribly esoteric approach were it not for the observation of
\cite{Bondal} that (at least for some varieties $X$) the derived
category $\Db(\Coh X)$ is equivalent to $\Db(Q\catrep)$ for some
quiver $Q$. This quiver just has to be the one of the low energy
effective field theory.

But the quiver $Q$ is not uniquely determined, and in general there
are different quivers $Q_i$ having equivalent derived categories
$\Db(Q_i\catrep)$. Can they all define the same low-energy physics? Of
course they should. Note that we no longer have to talk about coherent
sheaves or \CYms, but can turn this into a pure gauge-theory
statement:
\begin{conjecture}[Berenstein--Douglas]
  Two quivers $Q_1$, $Q_2$ are Seiberg dual if and only if
  $\Db(Q_1\catrep) \simeq \Db(Q_2\catrep)$.
\end{conjecture}
Please note that there is a lot more information in this conjecture
than a prescription like ``flip some arrows and fix up the
superpotential''. There is no ambiguity, no guesswork involved. The
appearance (or not) of superpotentials is completely forced on you.

This is very easy to state, but not at all obvious how those quivers
look like in general. It is not even clear that what you would naively
write down as BDS-dual quivers have equivalent derived representation
categories. I will explore this in the following.

\section{The Berenstein--Douglas Toy Quiver}

The only example for a pair of BDS-dual quivers
in~\cite{BerensteinDouglas} is the following:
\begin{equation}
\label{eq:BDtoyQ}
\begin{split}
  Q_1 \eqdef&\quad
  \xymatrix{
    \qnode{1} \ar[r]^\alpha & 
    \qnode{2} \ar[r]^\beta  & 
    \qnode{3}
  }
  \\
  Q_2 \eqdef&\quad
  \xymatrix{ 
    \qnode{1} & 
    \qnode{2} \ar[l]_{\phi_{21}} & 
    \qnode{3} \ar[l]_{\phi_{32}}
  }
  ~ \phi_{32}\phi_{21}  =0
\end{split}
\end{equation}
There is no oriented cycle, so one cannot write down a gauge invariant
superpotential. But there is one relation in $Q_2$ which should derive
from a superpotential. Obviously there is something missing, which I
will explain in section~\ref{sect:physQuiv}.

For now it does serve as a very nice toy model where we can try to
understand the equivalence of derived categories, see how the
computations work and fix conventions. I will work purely in the
derived category (which is more general and in some sense simpler), as
opposed to the module theoretic approach of~\cite{BerensteinDouglas}.

\subsection{Path Algebra, Representations and Modules}

When writing down the superpotential constraints we always use an
algebraic notation like
$\left<\text{arrow}\,1\right>\left<\text{arrow}\,2\right>
=0$. Formalizing this leads to the following:
\begin{definition}
  The \textdef{path algebra} $\C{} Q$ of a quiver $Q$ is the
  algebra (over $\C{}$) generated by the arrows of the quiver, and
  subject to the relations explained below.
  Here ``arrows'' stands for two kinds of arrows:
  \begin{itemize}
  \item The arrows $\phi_{ij}$ in the quiver, going from node $i$ to
  node $j$.
  \item For each node $i$ include a ``zero length'' arrow $e_i$, not
  drawn in the quiver diagram.
  \end{itemize}  
  The relations between the generators of the path algebra are the
  relations in the quiver. In addition to that we demand that every
  product is zero unless the arrows fit together, i.e. the only
  nonzero products are $\phi_{ij}\phi_{jk}$, $e_i \phi_{ij}\eqdef
  \phi_{ij}$, $\phi_{ij} e_j \eqdef \phi_{ij}$, $e_i^2=e_i$.
\end{definition}
Note that the composition of arrows is in the ``intuitive'' order,
which is \emph{opposite} to the composition of functions: If you
have $f:X\to Y$ and $g:Y\to Z$ then usually their composition is
denoted $g \circ f$. This means that in the representation of a
quiver, the vectors attached to the nodes should be thought of as row
vectors, and the matrices act by right-multiplication.
\begin{example}
  Take the quiver $Q=Q_1$ of eq.~\eqref{eq:BDtoyQ}. Then $\C{}Q$ is
  $6$-dimensional as a vector space:
  \begin{equation}
    \C{}Q = \Span_\C{}\Big( 
      e_1, e_2, e_3, 
      \alpha, \beta, \alpha \beta
    \Big)
  \end{equation}
\end{example}
\textbf{What is it good for?} Introducing the path algebra seems to be
completely formal. Although not obvious, it does simplify the
following computations considerably. The reason for this is the
following elementary fact:
\begin{theorem}
  Let $Q$ be an arbitrary quiver. Then the category of representations
  is the same as the category of $\C{}Q$-modules:
  \begin{equation}
    Q\catrep ~=~ \C{}Q\catmod
  \end{equation}
\end{theorem}
So instead of dealing with representations we can work with
modules. This is good because we have the following class of
manageable modules:
\begin{example}
  Let $P_i \eqdef \big( \C{}Q \big) e_i$, the paths ending at node
  $i$. Then $P_i$ is a $\C{}Q$ left module in the obvious sense:
  \begin{equation}
    \phi \cdot (p e_i) = (\phi p) e_i 
    \quad \forall \phi,p \in \C{}Q
  \end{equation}
\end{example}
If you have a path from node $i$ to $j$, then you can multiply $P_i$
on the right and land in $P_j$. This is a $\C{}Q$ module homomorphism,
and moreover all such homomorphisms come from such paths:
\begin{theorem}
  $  
    \dim_\C{} \Hom_{\C{}Q\catmod}(P_i, P_j) = 
    \# \big\{ \text{\rm Independent paths}~i\to j \big\}
  $
\end{theorem}
Of course there are other $\C{}Q$-modules. But they are not important
because of the following fact:
\begin{theorem}
  \label{thm:projective}
  The $P_i$ are projective $\C{}Q$-modules, and every projective
  module is a direct sum of $P_i$'s.
\end{theorem}
In the derived category $\Db(\C{}Q\catmod)$ every object is isomorphic
to its projective resolution, so it suffices to know just the
projective modules.

\subsection{The Derived Category}

So what is $\Db(\C{}Q\catmod)$, the derived category of
$\C{}Q$-modules? Well to define the category you need the objects and
the morphisms. The first is the easy part:
\begin{definition}
  The objects of $\Db(\C{}Q\catmod)$ are bounded complexes of
  $\C{}Q$-modules, i.e. a chain of modules and module homomorphisms
  \begin{equation}
    \cdots 
    \stackrel{d_{n-2}}{\longrightarrow}
    M_{n-1} 
    \stackrel{d_{n-1}}{\longrightarrow}
    M_{n} 
    \stackrel{d_{n}}{\longrightarrow}
    M_{n+1} 
    \stackrel{d_{n+1}}{\longrightarrow}
    \cdots
  \end{equation}
  such that going twice is zero, and only finitely many $M_n$ are
  nonzero.
\end{definition}
Now there is an obvious notion of ``map'' from one complex
$f:M_\bullet\to N_\bullet$, given by maps of the modules such that
\begin{equation}
  \xymatrix{
    \cdots  \ar[r] & 
    M_{n-1} \ar[r] \ar[d]_{f_{n-1}} & 
    M_{n}   \ar[r] \ar[d]_{f_{n}} & 
    M_{n+1} \ar[r] \ar[d]_{f_{n+1}} & 
    \cdots 
    \\
    \cdots  \ar[r] & 
    N_{n-1} \ar[r] & 
    N_{n}   \ar[r] & 
    N_{n+1} \ar[r] & 
    \cdots 
  }
\end{equation}
commutes. Those chain maps are \emph{not} the morphisms in the derived
category. The derived category contains less information than the
category of chain complexes. We get the morphisms of the derived
category if we apply the following to the set of chain maps:
\begin{enumerate}
\item ``Invert quasi-isomorphisms'' \\
  Maybe there is no (nonzero) chain map $M_\bullet\to N_\bullet$, but
  there exists another complex $M_\bullet'$ with the same homology and
  a nonzero chain map $f':M_\bullet'\to N_\bullet$. In the derived
  category you include $f'$ in the morphisms from $M$ to $N$.

  This sounds horribly complicated, but if all modules in the complex
  are already projective then this cannot happen. This is the reason
  for working only with the modules $P_i$.
\item ``Go to the homotopy category of chain complexes'' \\
  There is an equivalence relation (homotopy) on the chain maps that
  you have to mod out to get the morphisms of the derived category.
  A chain map $f:M_\bullet\to N_\bullet$ is homotopic to the zero map
  if there exists maps $s_n:M_{n}\to N_{n-1}$ such that
  \begin{equation}
    f_n = s_{n+1} d_{n} + d_{n-1} s_{n}
  \end{equation}
  In pictures:
  \begin{equation}
    \vcenter{\xymatrix{
      \strut  \cdots  \ar[r] & 
      M_{n-1} \ar[r] \ar[d]|{f_{n-1}} \ar@{.>}[dl]|{s_{n-1}} & 
      M_{n}   \ar[r] \ar[d]|{\vphantom{\displaystyle M}f_{n}}   
                                      \ar@{.>}[dl]|{s_{n}}   & 
      M_{n+1} \ar[r] \ar[d]|{f_{n+1}} \ar@{.>}[dl]|{s_{n+1}} & 
      \strut \cdots                   \ar@{.>}[dl]|{s_{n+2}}
    \\
      \strut \cdots  \ar[r] & 
      N_{n-1} \ar[r] & 
      N_{n}   \ar[r] & 
      N_{n+1} \ar[r] & 
      \strut \cdots 
    }}
  \end{equation}  
\end{enumerate}

\subsection{Computations in the Derived Category}
\label{sect:ComputationEndBD}

%Because we invert the quasiisomorphisms we can just replace every
%object in the derived category by its projective resolution, so we can
%assume that every module is projective.

Enough of the theory, lets try to see how this works (we'll need the
results of this section later). Fix $Q\eqdef Q_1$ of
eq.~\eqref{eq:BDtoyQ}, and consider the following $3$ objects of
$\Db(\C{}Q\catmod)$: 
\begin{equation}
  \label{eq:T123defBD}
  \begin{split}
      T_1 \eqdef \quad
      \xymatrix@!{
        0 \ar[r] & 0 \ar[r] & \underline{P_1} \ar[r] & 0 
      }
    \\
      T_2 \eqdef \quad
      \xymatrix@!{
        0 \ar[r] & 0 \ar[r] & \underline{P_3} \ar[r] & 0 
      }
    \\
      T_3 \eqdef \quad
      \xymatrix@!{
        0 \ar[r] & P_2 \ar[r]|\beta & \underline{P_3} \ar[r] & 0 
      }
  \end{split}
\end{equation}
where the nonzero map in $T_3$ is right-multiplication by $\beta$, in
abuse of notation again called $\beta$. The underlined entry denotes
the one at position $0$.

I want to identify the morphisms $\Hom(T_a\T{k},T_b) \eqdef
\Hom_{\Db(\C{}Q\catmod)}(T_a\T{k},T_b)$ in the derived category
($T_a\T{k}$ is $T_a$ shifted by $k$ positions to the left). Here are
some examples, the computation is always along these lines:
\begin{descriptionlist}
\item[$\Hom(T_2,T_1)=0$]
  because there is no module-homomorphism $P_2\to P_1$ (there is no
  path in the quiver from node $2$ to $1$).
\item[$\Hom(T_1,T_1)=\Span(e_1)\simeq \C{}$] because the only
  module-homomorphism $P_1\to P_1$ is multiplication with a constant.
  Or in other words, $e_1$ is the only path from node $1$ to itself.
  Note that --- again by abuse of notation --- I also denote the
  morphism in the derived category by $e_1$.
\item[$\Hom(T_1,T_2)=\Span(\alpha \beta) \simeq \C{}$] by the same
  reason as above; The map cannot be homotopic to the zero map because
  too many entries are zero.
\item[$\Hom(T_2,T_3)\simeq \C{}$] coming from multiplying $P_3$ by a
  constant. There is no homomorphism $P_3\to P_2$ so there is no
  nontrivial chain homotopy.
\item[$\Hom(T_1,T_3)=0$] because the possible chain map is homotopic
  to zero:
  \begin{equation}
    \label{eq:T1T3homotopy}
    \vcenter{\xymatrix{
      \strut  
      0   \ar[r] \ar[d] & 
      0   \ar[r] \ar[d]|{\vphantom{\displaystyle Mg}0} 
                                      \ar@{.>}[dl]|{0}     & 
      \underline{P_1}
          \ar[r] \ar[d]|{\vphantom{\displaystyle Mg}\alpha\beta}   
                                      \ar@{.>}[dl]|{\alpha}   & 
      0 \strut \ar@{.>}[dl]|{0} \ar[d] 
    \\
      \strut 
      0    \ar[r] & 
      P_2  \ar[r]|{\beta} & 
      \underline{P_3}  \ar[r] & 
      0
    }}
  \end{equation}
\item[$\Hom(T_1\T{1},T_3)=0$] since --- although there would be a
  nontrivial maps of the modules $\alpha:P_1\to P_2$ --- there are no
  chain maps:
  \begin{equation}
    \vcenter{\xymatrix{
      \strut  
      0   \ar[r]      & 
      P_1 \ar[r] \ar[d]|{\vphantom{\displaystyle Mg}\alpha} &
      \underline{0}
                 \ar[r] \ar[d]|{\vphantom{\displaystyle Mg}0} 
                 \ar@{}[dl]|{\displaystyle \lightning}  &
      0 
    \\
      \strut 
      0    \ar[r] & 
      P_2  \ar[r]|{\beta} & 
      \underline{P_3}  \ar[r] & 
      0
    }}
   \qquad
   \alpha \beta \not= 0
  \end{equation}      
\end{descriptionlist}
Along these lines one can determine all $\Hom$'s. The result is that
\begin{equation}
  \label{eq:HomTiT}
  \Hom(T_a\T{i},T_b)=0 \qquad \text{unless}~i=0
\end{equation}
so the only nontrivial morphisms are between the unshifted
$T$'s. 
\begin{table}[htbp]
  \centering
  \begin{tabular}{c|ccc}
    $A\backslash B$ & $T_1$ & $T_2$ & $T_3$ \\ \hline
    $T_1$ &
      $[0,e_1]$ & 
      $[0,\alpha\beta]$ &
      % none
    \\ 
    $T_2$ &
      & % none
      $[0,e_3]$ & 
      $[0,e_3]$
    \\ 
    $T_3$ &  
      & % none
      & % none
      $[e_2,e_3]$ 
  \end{tabular}
  \caption{Summary of $\Hom(A,B)$ in $\Db(\C{}Q\catmod)$.}
  \label{tab:BDtoyHom}
\end{table}
Since the maximum width of the complexes is $2$ every derived
morphism is generated by a pair of module maps, which I denote $[f_1,
f_2]$ in table~\ref{tab:BDtoyHom}.

A point worth noting is that although there are nonzero maps $T_1\to
T_2$ and $T_2\to T_3$, the composition is zero in the derived category
since $\Hom(T_1,T_3)=0$. Although neither individual map is homotopic
to zero, their composition is, see eq.~\eqref{eq:T1T3homotopy}.

\subsection{Fourier--Mukai Transformation}
\label{sect:Trans}

Checking the equivalence of the derived categories
$\Db(\C{}Q_1\catmod)$ and $\Db(\C{}Q_2\catmod)$ just from the
definition certainly would be a formidable task. However there is some
more machinery that makes this actually possible, Fourier--Mukai
transformations. The single most useful thing about the whole derived
category business is that Fourier--Mukai induces equivalences of
derived categories (see e.g.~\cite{BridgelandKingReid}).

Now for the derived categories of quiver path algebra modules this
boils down to something quite manageable: If you have an element $T\in
\Db(\C{}Q_1\catmod)$ (i.e. a complex of $\C{}Q_1$-modules) then you
get an algebra $\Hom(T,T)\eqdef \End(T)$. If this algebra is
isomorphic to $\C{}Q_2$ and if $T$ has some nice properties then
$\C{}Q_1\catmod$ and $\C{}Q_1\catmod$ are derived equivalent.

To be precise~\cite{Rickard} showed the following:
\begin{definition}
  $T\in \Db(\C{}Q\catmod)$ is called a \textdef{tilting complex} if
  \begin{enumerate}
  \item $\Hom(T\T{i},T)=0$ for $i\not=0$
  \item Summands of direct sums of copies of $T$ generate
  $\Db(\C{}Q\catmod)$.
  \end{enumerate}
\end{definition}
\begin{theorem}
  \label{thm:tilting}
  The derived categories $\Db(\C{}Q_1\catmod)$ and
  $\Db(\C{}Q_2\catmod)$ are equivalent if and only if there exists a
  tilting complex $T\in \Db(\C{}Q_1\catmod)$ such that $\End(T)\simeq
  \C{}Q_2$. 
\end{theorem}
So in our case (the quivers in eq.~\eqref{eq:BDtoyQ}) set $T=T_1\oplus
T_2 \oplus T_3 \in \Db(\C{}Q_1\catmod)$ with the $T_i$ defined in
eq.~\eqref{eq:T123defBD}.

I claim that $T$ is a tilting complex: There are no $\Hom$'s between
$T\T{i}$ and $T$ because of eq.~\eqref{eq:HomTiT}, it remains to show
that one can generate $\Db(\C{}Q_1\catmod)$ from $T_1$, $T_2$
and~$T_3$. For that it suffices to generate the stalk complexes for
$P_1$, $P_2$ and~$P_3$, i.e. complexes of the form $0\to
\underline{P_i}\to 0$ (by abuse of notation again denoted $P_i$).

The only nontrivial task is to generate the stalk complex of $P_2$.
Here ``generate'' means to generate as derived category, so it
includes operations like the shift and mapping cones. Let $f=[0,e_3]$
the generator of $\Hom(T_2,T_3)$, then by definition the mapping cone
is
\begin{equation}
  M(f) = \quad
  \xymatrix{
    0 \ar[r] & 
    P_2 \oplus P_3 \ar[r]^-{\left(%
      \begin{smallmatrix}
        \beta \\ e_3
      \end{smallmatrix} \right) }&
    \underline{P_3} \ar[r] & 0 
  }
\end{equation}
Now in the derived category $M(f)\simeq P_2\T{1}$. To see this check
that the compositions of the obvious chain maps $M(f)
\begin{smallmatrix} \rightarrow \\ \leftarrow \end{smallmatrix}
P_2\T{1}$ are homotopic to the identity:
\begin{equation}
  \vcenter{
  \xymatrix{
      0 \ar[r] 
    & 
      P_2 \oplus P_3 
      \ar[r]^-{\left(%
        \begin{smallmatrix}
          \beta \\ 1
        \end{smallmatrix} \right) 
      }
      \ar[d]_-{\left(%
        \begin{smallmatrix}
          1 \\ 0
        \end{smallmatrix} \right) 
      }
    &
      \underline{P_3} \ar[r] 
      \ar[d]
      \ar@{.>}[ddl]|(0.3){\left(%
        \begin{smallmatrix}
          0 & 1
        \end{smallmatrix} \right) 
      }
    & 0 
    & 
      0 \ar[r] 
    & 
      P_2  \ar[r]
      \ar[d]_-{\left(%
        \begin{smallmatrix}
          1 & -\beta
        \end{smallmatrix} \right) 
      }
    &
      \underline{0} \ar[r] 
      \ar[d]
    & 0 
    \\
      0 \ar[r] 
    & 
      P_2  \ar[r]
      \ar[d]_-{\left(%
        \begin{smallmatrix}
          1 & -\beta
        \end{smallmatrix} \right) 
      }
    &
      \underline{0} \ar[r] 
      \ar[d]
    & 0 
    &
      0 \ar[r] 
    & 
      P_2 \oplus P_3 
      \ar[d]_-{\left(%
        \begin{smallmatrix}
          1 \\ 0
        \end{smallmatrix} \right) 
      }
      \ar[r]^-{\left(%
        \begin{smallmatrix}
          \beta \\ 1
        \end{smallmatrix} \right) 
      }
    &
      \underline{P_3} \ar[r] 
      \ar[d]
    & 0 
    \\
      0 \ar[r] 
    & 
      P_2 \oplus P_3 
      \ar[r]^-{\left(%
        \begin{smallmatrix}
          \beta \\ 1
        \end{smallmatrix} \right) 
      }
    &
      \underline{P_3} \ar[r] 
    & 0 
    & 
      0 \ar[r] 
    & 
      P_2  \ar[r]
    &
      \underline{0} \ar[r] 
    & 0
    \\
  }}
\end{equation}
Thus $T=T_1\oplus T_2 \oplus T_3$ is a tilting complex. We determined
$\End(T)$ in section~\ref{sect:ComputationEndBD}, it is the path
algebra of $Q_2$. The $3$ orthogonal projectors correspond to the
zero length arrows at the nodes, and the $2$ remaining endomorphisms
are just the arrows between the nodes (satisfying the relation that
their composition is $0$). With other words $\End(T)\simeq \C{}Q_2$
and therefore $\Db(\C{}Q_1\catmod)\simeq\Db(\C{}Q_2\catmod)$.

\section{Physical Quivers}
\label{sect:physQuiv}

\subsection{Quivers with Superpotential}

It is not true that every quiver corresponds to an $\Ncal=1$ gauge
theory, only the subset of quivers where the relations are derived
from a superpotential do. 
\begin{definition}
  A \textdef{superpotential} $W$ (for some oriented graph $\Gamma$) is
  the trace over a linear combination of oriented cycles, i.e.
  \begin{equation}
    W \in \tr \bigoplus e_i \, (\C{}\Gamma)\, e_i
  \end{equation}
\end{definition}
Of course at the level of the path algebra, the trace is just a formal
function with the cyclic permutation property.

Then a $\Ncal=1$ gauge theory corresponds to
\begin{definition}
  A \textdef{quiver with superpotential} $W$ is a quiver such that the
  relations are\footnote{The authors conventions are that derivations
  act from the left} $\frac{\partial W}{\partial \alpha_i}=0$, where
  $\alpha_i$ are the arrows of the quiver.
\end{definition}

\subsection{A Sample Duality Pair}

As already mentioned the BDS-dual pair in eq.~\eqref{eq:BDtoyQ} is
not a pair of ``quivers with superpotential''. The obvious thing to do
is to close up $Q_2$ with a third arrow. But this changes the derived
category and $\Db(Q_1\catrep) \not\simeq \Db(Q_2'\catrep)$, as I will
demonstrate in section~\ref{sect:gldim}.

The simplest example for actual BDS-dual quivers with superpotential
is the following:
\begin{equation}
\label{eq:QuivEx}
\begin{split}
  Q_1 \eqdef&\quad
  \vcenter{\xymatrix{ 
      \qnode{1} \ar[d]_{\alpha} 
    \\
      \qnode{2} \ar[d]_{\beta} 
    &
      \qnode{4} \ar[l]_{\delta}
    \\
      \qnode{3} \ar@(l,l)[uu]_{\gamma} 
  }}
  \begin{array}{c}
    \alpha \beta  = 0 \\ 
    \beta  \gamma = 0 \\ 
    \gamma \alpha = 0 
  \end{array}
  \quad
  W_1= \tr \big( \alpha \beta  \gamma \big)
  \\
  Q_2 \eqdef&\quad
  \vcenter{\xymatrix{ 
      \qnode{1} 
    \\
      \qnode{2} \ar[u]^{\phi_{21}}  \ar[r]^{\phi_{24}} 
    &
      \qnode{4} \ar[dl]^{\phi_{43}}
    \\
      \qnode{3} \ar[u]^{\phi_{32}} 
  }}
  \begin{array}{c}
    \phi_{32}\phi_{24}=0 \\
    \phi_{24}\phi_{43}=0 \\
    \phi_{43}\phi_{32}=0
  \end{array}
  \quad
  W_2= \tr\big( \phi_{32} \phi_{24} \phi_{43} \big)
\end{split}
\end{equation}
Note that the quivers can be distinguished by the direction of the
arrow not in the cycle. The remainder of this section will be devoted
to proving their derived equivalence, of course by using
theorem~\ref{thm:tilting}.

\subsection*{The Tilting Complex}

I claim that $T\eqdef T_1\oplus T_2\oplus T_3\oplus T_4 \in
\Db(\C{}Q_1\catmod)$ is a tilting complex, where
\begin{equation}
  \label{eq:T1234defV}
  \begin{split}
      T_1 \eqdef \quad
      \xymatrix@=21mm@!0{
        0 \ar[r] & \underline{P_1} \ar[r] & 0 \ar[r] & 0 
      }
    \\
      T_2 \eqdef \quad
      \xymatrix@=21mm@!0{
        0 \ar[r] & \underline{P_1 \oplus P_4} 
        \ar[r]|-{
          \left(
          \begin{smallmatrix}
            \alpha \\ \delta
          \end{smallmatrix}
          \right)} & 
        P_2 \ar[r] & 0 
      }
    \\
      T_3 \eqdef \quad
      \xymatrix@=21mm@!0{
        0 \ar[r] & \underline{P_3} \ar[r] & 0 \ar[r] & 0 
      }
    \\
      T_4 \eqdef \quad
      \xymatrix@=21mm@!0{
        0 \ar[r] & \underline{P_4} \ar[r] & 0 \ar[r] & 0 
      }
    \\
  \end{split}
\end{equation}
First we need to show that $\Hom(T\T{k},T)=0~\forall i\not=0$. This is
straightforward, I discuss the two prototypical sample cases:
\begin{descriptionlist}
\item[$\Hom(T_1\T{-1},T_2)=0$] since the chain map is homotopic to
  zero:
  \begin{equation}
    \label{eq:T1T3homotopyV}
    \vcenter{\xymatrix{
      \strut  
      0   \ar[r] \ar[d] & 
      \underline{0}   
      \ar[r] \ar[d]|{\vphantom{\displaystyle Mg}0} 
                                      \ar@{.>}[dl]|{0}     & 
      P_1
      \ar[r] \ar[d]|{\vphantom{\displaystyle Mg}\alpha}   
                                      \ar@{.>}[dl]|{\left(
                                        \begin{smallmatrix}
                                          1 & 0
                                        \end{smallmatrix}\right)} & 
      0 \strut \ar@{.>}[dl]|{0} \ar[d] 
    \\
      \strut 
      0    \ar[r] & 
      \underline{P_1 \oplus P_4} 
        \ar[r]_-{
          \left(
          \begin{smallmatrix}
            \alpha \\ \delta
          \end{smallmatrix}
          \right)} & 
      \underline{P_2}  \ar[r] & 
      0
    }}
  \end{equation}
\item[$\Hom(T_2\T{1},T_3)=0$] since there is no chain map:
  \begin{equation}
    \vcenter{\xymatrix{
      \strut  
      0   \ar[r]      & 
      P_1\oplus P_4         
        \ar[r]^-{
          \left(\begin{smallmatrix}
            \alpha \\ \delta
          \end{smallmatrix}\right)} 
        \ar[d] &
      \underline{P_2}
                 \ar[r] \ar[d]|{\vphantom{\displaystyle Mg}\beta} 
                 \ar@{}[dl]|{\displaystyle \lightning}  &
      0 
    \\
      \strut 
      0    \ar[r] & 
      0    \ar[r] & 
      \underline{P_3}  \ar[r] & 
      0
    }}
   \qquad
     \left(\begin{smallmatrix} \alpha \\ \delta \end{smallmatrix}\right)
     \beta 
   =
     \left(\begin{smallmatrix} 0 \\ \delta \beta \end{smallmatrix}\right)
   \not=
     \left(\begin{smallmatrix} 0 \\ 0 \end{smallmatrix}\right)
  \end{equation}      
\end{descriptionlist}
Second we need to check that we can generate the whole
$\Db(\C{}Q_1\catmod)$. As in section~\ref{sect:Trans} the only
possible problem is to generate the stalk complex of $P_2$. But we
can get this as the mapping cone of the obvious map $f:T_2\to
T_1\oplus T_4$:
\begin{equation}
  \begin{split}    
    M(f) =&\quad
    \xymatrix{
      0 \ar[r] & 
      P_1 \oplus P_4 \ar[r]^-{\left(%
        \begin{smallmatrix}
          \alpha & 1 & 0 \\ 
          \delta & 0 & 1
        \end{smallmatrix} \right) }&
      \underline{P_2\oplus P_1\oplus P_4} \ar[r] & 0 
    }
  \\ 
    =&\quad
    \xymatrix{
      0   \ar[r] & 
      \underline{P_2} \ar[r] &       
      0          
    }    
  \end{split}
\end{equation}

\subsection*{The Dual Quiver}

We have to determine $\End(T)$. The nonvanishing morphisms between the
$T_i$ are listed in table~\ref{tab:HomV} using the same notation as in
table~\ref{tab:BDtoyHom}.
\begin{table}[htbp]
  \centering
  \begin{tabular}{c|cccc}
    $A\backslash B$ & $T_1$ & $T_2$ & $T_3$ & $T_4$ \\ \hline
    $T_1$ &
      $\left[1,0 \right]$ & 
      & % none $\left[,\right]$ &
      & % none $\left[,\right]$ &
      % none $\left[,\right]$ 
    \\ 
    $T_2$ &
      $\left[ \left(\begin{smallmatrix} 1 \\ 0
              \end{smallmatrix}\right)
              ,0 \right]$ &
      $\left[ \left(\begin{smallmatrix} 1 & 0 \\ 0 & 1
                    \end{smallmatrix}\right)
              ,1 \right]$ & 
      & % none $\left[,\right]$ &
      $\left[ \left(\begin{smallmatrix} 0 \\ 1
                    \end{smallmatrix}\right)
              ,0 \right]$ 
    \\ 
    $T_3$ &
      $\left[\gamma,0 \right]$ &
      $\left[ \left(\begin{smallmatrix} \gamma & 0
                    \end{smallmatrix}\right)
              ,0 \right]$ & 
      $\left[1,0 \right]$ &
      % none $\left[,\right]$ 
    \\ 
    $T_4$ &
      & % none $\left[,\right]$ &
      & % none $\left[,\right]$ &
      $\left[\delta\beta,0 \right]$ &
      $\left[1,0 \right]$ 
  \end{tabular}
  \caption{Summary of $\Hom(A,B)$ in $\Db(\C{}Q_1\catmod)$.}
  \label{tab:HomV}
\end{table}
Let $\phi_{ab}\in \Hom(T_a,T_b)$ be the generator, then they satisfy
the following relations:
\begin{descriptionlist}
\item[$\phi_{31}=\phi_{32}\phi_{21}$] by matrix multiplication.
\item[$\phi_{32}\phi_{24}=0$] by matrix multiplication.
\item[$\phi_{24}\phi_{43}=0$] since the composition is in
  $\Hom(T_2,T_3)=0$, the chain map is homotopic to zero.
\item[$\phi_{43}\phi_{32}=0$] since $\delta \beta \gamma=0$.
\end{descriptionlist}
After eliminating $\phi_{31}$ in the endomorphism algebra via the
first relation we see that $\End(T)=\C{}Q_2$ of eq.~\eqref{eq:QuivEx}.

\subsection{Beyond Tilting Modules}

In the summand $T_2$ of the tilting complex eq.~\eqref{eq:T1234defV}
the map $\left(\begin{smallmatrix} \alpha \\ \delta
  \end{smallmatrix}\right) P_1 \oplus P_4\to P_2$ is neither injective
nor surjective, e.g. $e_2\in P_2$ is not in the image and it maps
$(\gamma,0)\in P_1\oplus P_4$ to zero. This means that the complex
$T_2$ has non-zero homology at positions $0$ and~$1$, so it is not the
stalk complex of some tilting module. Although this does not prove
that there might not be some other tilting module it illustrates that
the derived category approach is more powerful than the one
of~\cite{BerensteinDouglas}.

In the case of toric dual quivers that I will treat next it will also
be the case that the tilting complex has homology in more than one
position.

However in all the examples that I will treat it suffices to use a
tilting complex that consists of the stalk complexes of the projective
modules, except one. Had we chosen to use only the $P_i$ stalk
complexes this would have also been a tilting complex, but the dual
quiver would be the original one. In this sense we always dualize only
a single node of the quiver, i.e. a single gauge group.

\section{Toric Duality is Seiberg Duality}

It has been argued that toric duality is Seiberg duality. This
provides us with more examples of dual quivers, provided that we
believe in the algorithm to read off the gauge theory data from the
toric variety.

\subsection{Review of Toric Duality}

Toric duality was suggested in~\cite{HananyHe1} as a new gauge theory
duality coming from different resolutions of a $\C{3}$-orbifold. 

\begin{figure}[!p]
  \centering
\begin{picture}(0,0)%
\includegraphics{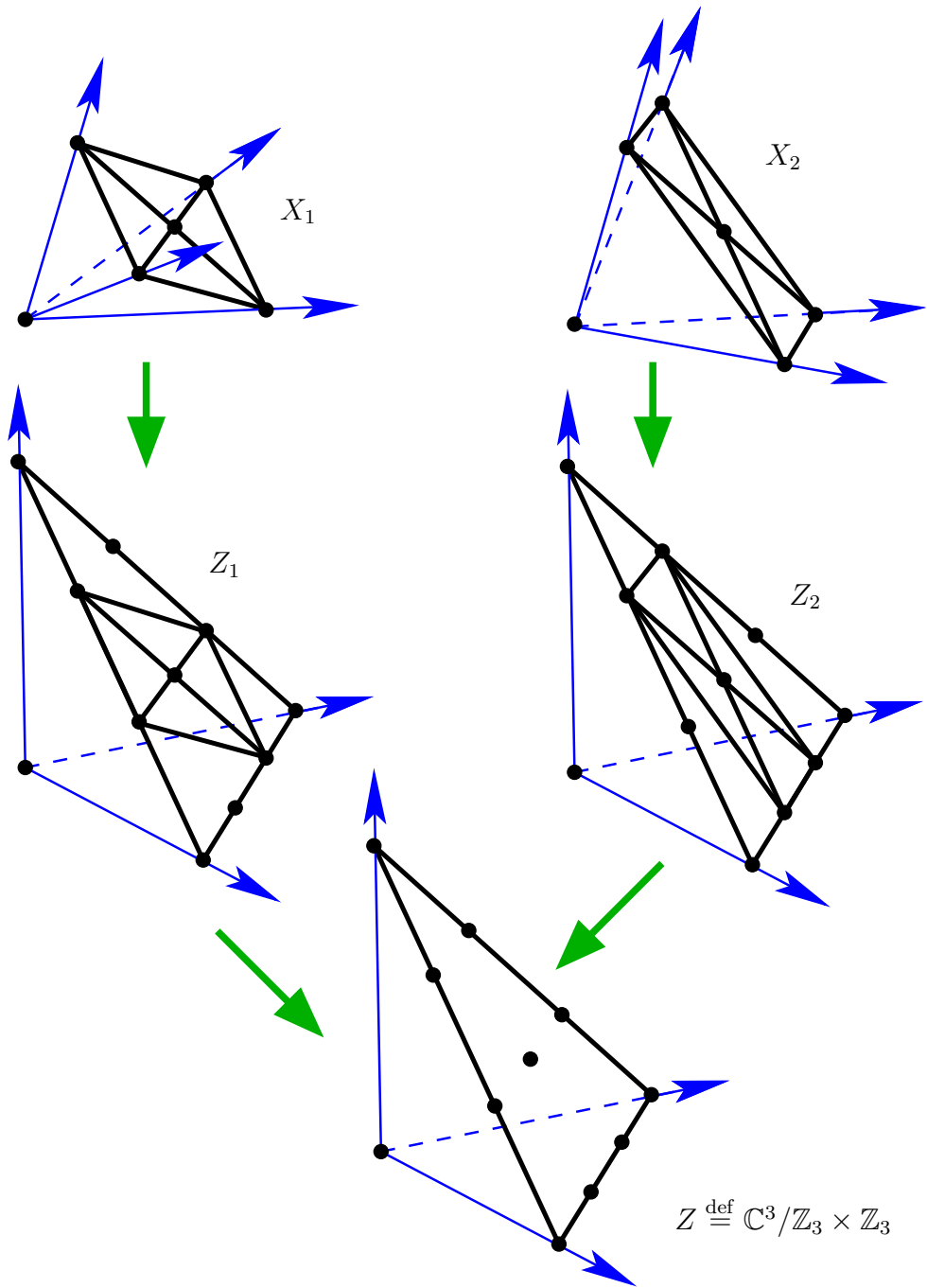}%
\end{picture}%
\setlength{\unitlength}{4144sp}%
\begingroup\makeatletter\ifx\SetFigFont\undefined%
\gdef\SetFigFont#1#2#3#4#5{%
  \reset@font\fontsize{#1}{#2pt}%
  \fontfamily{#3}\fontseries{#4}\fontshape{#5}%
  \selectfont}%
\fi\endgroup%
\begin{picture}(5842,8148)(15,-8858)
\put(4816,-1740){\makebox(0,0)[lb]{\smash{\SetFigFont{12}{14.4}{\rmdefault}{\mddefault}{\updefault}{$X_2$}%
}}}
\put(4967,-4519){\makebox(0,0)[lb]{\smash{\SetFigFont{12}{14.4}{\rmdefault}{\mddefault}{\updefault}{$Z_2$}%
}}}
\put(4245,-8475){\makebox(0,0)[lb]{\smash{\SetFigFont{12}{14.4}{\rmdefault}{\mddefault}{\updefault}{$Z\eqdef \C{3}/\Z{3}\times \Z{3}$}%
}}}
\put(1749,-2083){\makebox(0,0)[lb]{\smash{\SetFigFont{12}{14.4}{\rmdefault}{\mddefault}{\updefault}{$X_1$}%
}}}
\put(1296,-4318){\makebox(0,0)[lb]{\smash{\SetFigFont{12}{14.4}{\rmdefault}{\mddefault}{\updefault}{$Z_1$}%
}}}
\end{picture}
  \caption{$X_1$ and~$X_2$ are subvarieties of $Z_1$ and~$Z_2$, 
           which are both resolutions of $Z$}
  \label{fig:toricdual}
\end{figure}

For concreteness consider the toric singularity $Z=\C{3}/Z_3\times
\Z{3}$ depicted in figure~\ref{fig:toricdual}.
\afterpage{\clearpage} 
Pick two different resolutions $Z_1$ and~$Z_2$.  Then a subvariety
(given by a subfan) $X_1\subset Z_1$ and $X_2\subset Z_2$ are what I
will call \textdef{weak toric dual}. By the inverse algorithm
of~\cite{HananyHe1} one can associate a gauge theory to $X_1$
and~$X_2$, and those are also called weak toric dual. Since $X_1$ and
$X_2$ do not really have anything in common, weak toric duality is not
terribly interesting.

The interesting case is when $X_1$ and~$X_2$ are both the same space
(a line bundle over a toric surface), but embedded in different
resolutions $Z_1$ and $Z_2$. Then the gauge theory obviously should be
the same, while one in general finds different quivers. I will call
this case \textdef{strong toric duality}, or just toric duality. Note
that this is the case depicted in figure~\ref{fig:toricdual}, the two
bases of the line bundle are isomorphic toric varieties (as was noted
in~\cite{HananyHe2}).

In~\cite{HananyHe3,Plesser} it was then argued that strong toric dual
quivers are Seiberg dual gauge theories (slightly different dualities
were considered in~\cite{Lust}). If you believe that the
quiver associated to the toric variety is the correct one then this
follows from the one-line argument that (see~\cite{BerensteinDouglas})
\begin{equation}
  \Db(\C{}Q_1\catmod)
  \simeq 
  \Db(\Coh X_1)
  =
  \Db(\Coh X_2)
  \simeq 
  \Db(\C{}Q_2\catmod)
\end{equation}
By the algorithm of~\cite{HananyHe1} (which I will not review here)
one can for example find toric dualities between the following
quivers:
% ================== begin big equation ========================
\savebox{\QuivToricA}{
\begin{math}
\begin{array}{c}
\displaystyle
  \vcenter{\xymatrix@R=1cm@C=0.433cm{ & 
      \qnode{3} 
      \ar@<-1ex>[ld]_{x_{7,8,9}}  %(0.3){x_7}_(0.5){x_8}_(0.7){x_9}  
      \ar[ld]
      \ar@<1ex>[ld]
    \\
      \qnode{1} 
      \ar@<-1ex>[rr]_{x_{1,2,3}}  %_(0.3){x_1}_(0.5){x_2}_(0.7){x_3}
      \ar[rr]
      \ar@<1ex>[rr]
    & & 
      \qnode{2} 
      \ar@<-1ex>[lu]_{x_{4,5,6}}  %_(0.3){x_4}_(0.5){x_5}_(0.7){x_6}
      \ar[lu]
      \ar@<1ex>[lu]
  }}
\\ % array linebreak
  \scriptstyle
    W = \tr\Big( 
    (x_2 x_6 - x_3 x_5) x_7 + \text{cycl.} \Big)
\end{array}
\end{math}
}
% ==============================================================
\savebox{\QuivToricB}{
\begin{math}
\begin{array}{c}
\displaystyle
  \vcenter{\xymatrix@R=1cm@C=1cm{ 
      \qnode{4} 
      \ar@<-0.5ex>[d]_{\delta_{0,1}}
      \ar@<0.5ex>[d]
    & 
      \qnode{3} 
      \ar@<-0.5ex>[l]_{\gamma_{0,1}}
      \ar@<0.5ex>[l]
    \\
      \qnode{1} 
      \ar@<-0.5ex>[r]_{\alpha_{0,1}}
      \ar@<0.5ex>[r]
    & 
      \qnode{2} 
      \ar@<-0.5ex>[u]_{\beta_{0,1}}
      \ar@<0.5ex>[u]
  }}
\\ % array linebreak
\scriptstyle
   W = \tr\big( 
      \alpha_0( \beta_0 \gamma_1 \delta_1 
              - \beta_1 \gamma_1 \delta_0 ) + ~
\hfill
\\ % array break
\hfill  
\scriptstyle
    + \alpha_1( \beta_1 \gamma_0 \delta_0 - \beta_0 \gamma_0 \delta_1 )
    \big) 
\end{array}
\end{math}
}
% ==============================================================
\savebox{\QuivToricC}{
\begin{math}
\begin{array}{c}
\displaystyle
  \vcenter{\xymatrix@R=1cm@C=1cm{ 
      \qnode{4} 
      \ar@<-0.5ex>[r]
      \ar@<0.5ex>[r]^{y_{9,10}}
    & 
      \qnode{3} 
      \ar@{}[dl]|{y_{5,6,11,12}}
      \ar@<-1.2ex>[dl]|(0.59){\phantom{Mg}}
      \ar@<-0.4ex>[dl]|(0.53){\phantom{Mg}}
      \ar@<00.4ex>[dl]|(0.47){\phantom{Mg}}
      \ar@<01.2ex>[dl]|(0.41){\phantom{Mg}}
    \\
      \qnode{1} 
      \ar@<-0.5ex>[r]_{y_{1,2}}
      \ar@<0.5ex>[r]
      \ar@<-0.5ex>[u]
      \ar@<0.5ex>[u]^{y_{7,8}}
    & 
      \qnode{2} 
      \ar@<-0.5ex>[u]_{y_{3,4}}
      \ar@<0.5ex>[u]
  }}
\\ % array linebreak
\scriptstyle
   W = \tr\big( 
      ~ y_{5}  ( y_{2} y_{4} - y_8 y_{10} ) -
\hfill
\\ % array break
\hfill  
\scriptstyle
      - y_{6}  ( y_{2} y_{3} - y_7 y_{10} ) - 
\\ % array break
\hfill  
\scriptstyle
      -  y_{11} ( y_{1} y_{4} - y_8 y_{9}  ) + 
\\ % array break
\hfill  
\scriptstyle
       + y_{12} ( y_{1} y_{3} - y_7 y_{9}  ) \big)
\end{array}
\end{math}
}
% ==============================================================
\begin{equation}
  \label{eq:Qtoricdual}
  \vcenter{\xymatrix@!C=35mm{
  & 
    \usebox{\QuivToricA}
%  \ar@(ul,ul)@2{<->}[dl]_{\txt{weak toric dual}}
  \ar@2{<->}[dl]_-{\txt{weak toric dual}}
  \\
    \usebox{\QuivToricB}
%  \ar@(d,d)@2{<->}[rr]_{\txt{strong toric dual}}
  \ar@2{<->}[rr]^{\txt{strong}}_{\txt{toric dual}}
  & & 
    \usebox{\QuivToricC}
%  \ar@(ur,ur)@2{<->}[ul]_{\txt{weak toric dual}}
  \ar@2{<->}[ul]_-{\txt{weak toric dual}}
  }}
\end{equation}
% ================== end big equation ==========================
In the following we will see (section~\ref{sec:ktheory}) that the
weakly dual quivers have non-equivalent derived categories, as
expected. In the rest of this section I will demonstrate that the
strong toric dual quivers above are BDS-dual.

\subsection{Toric Duality by Tilting}
\label{sec:torictilt}

It is a nice consistency check to actually show that the strong toric
dual quivers of eq.~\eqref{eq:Qtoricdual} are BDS-dual, as I will do
now. The starting point is the following quiver (the left one in
eq.~\eqref{eq:Qtoricdual}) with the relations derived from the
superpotential: 
\begin{equation}
  Q\eqdef \quad
  \vcenter{\xymatrix@R=15mm@C=15mm{ 
      \qnode{4} 
      \ar@<-0.7ex>[d]_{\delta_{0,1}}
      \ar@<0.7ex>[d]
    & 
      \qnode{3} 
      \ar@<-0.7ex>[l]_{\gamma_{0,1}}
      \ar@<0.7ex>[l]
    \\
      \qnode{1} 
      \ar@<-0.7ex>[r]_{\alpha_{0,1}}
      \ar@<0.7ex>[r]
    & 
      \qnode{2} 
      \ar@<-0.7ex>[u]_{\beta_{0,1}}
      \ar@<0.7ex>[u]
  }}
  \quad
  \begin{array}{c}
    \beta_0 \gamma_1 \delta_1  = \beta_1 \gamma_1 \delta_0  \\
    \gamma_0 \delta_1 \alpha_1 = \gamma_1 \delta_1 \alpha_0 \\
    \delta_0 \alpha_1 \beta_1  = \delta_1 \alpha_1 \beta_0  \\
    \alpha_0 \beta_1 \gamma_1  = \alpha_1 \beta_1 \gamma_0 
  \end{array}
  \quad
  \begin{array}{c}
    \beta_1 \gamma_0 \delta_0  = \beta_0 \gamma_0 \delta_1  \\
    \gamma_1 \delta_0 \alpha_0 = \gamma_0 \delta_0 \alpha_1 \\
    \delta_1 \alpha_0 \beta_0  = \delta_0 \alpha_0 \beta_1  \\
    \alpha_1 \beta_0 \gamma_0  = \alpha_0 \beta_0 \gamma_1 
  \end{array}
\end{equation}
Technically the path algebra is much harder to analyze since there are
arbitrary long paths, i.e. $\C{}Q$ is an infinite-dimensional algebra.

Since we cannot simply enumerate all paths I will first describe some
of the structure of the path algebra. First note that the relations
allow us to sort the indices at odd and at even positions (e.g.
counting the individual arrows from left to right starting at 1), not
more and not less. This implies the
\begin{definition}[Normal form for paths in $\C{}Q$]
  Consider the relation $\succ$ that compares the indices of the
  arrows of the quiver:
  \begin{equation}
    v \succ w 
    ~\stackrel{\text def}{\Longleftrightarrow}~
    v \in \big\{\alpha_1, \beta_1, \gamma_1, \delta_1\big\}
    ~\text{and}~
    w \in \big\{\alpha_0, \beta_0, \gamma_0, \delta_0\big\}
  \end{equation}
  and let $\preceq$ be the complement (comparing indices via $\leq$).
  Then for every path
  \begin{equation}
    p = p_1 p_2 \cdots p_k \in \C{}Q
  \end{equation}
  there exists a unique representative of the form $\tilde{p} =
  \tilde{p}_1 \tilde{p}_2 \cdots \tilde{p}_k$ such that the even and
  odd $\tilde{p}$'s are $\preceq$-ordered:
  \begin{equation}
    \tilde{p}_{2i} \preceq \tilde{p}_{2i+2}
    \quad\text{and}\quad
    \tilde{p}_{2i+1} \preceq \tilde{p}_{2i+3}
  \end{equation}
\end{definition}
The normal form implies another useful observation:
\begin{lemma}
  \label{lemma:grading}
  The path algebra $\C{}Q$ is $\Z{}\oplus\Z{}$-graded via
  \begin{equation}
    \mathop{grade}(p) = \Big(
      \text{length of}~p,~
      \text{sum of the indices of the individual arrows}
    \Big)
  \end{equation}
\end{lemma}
This is useful because we can sort the paths, for example first sort
by length and then break ties by comparing the sum of the
indices. Then we can understand the following:
\begin{lemma}
  \label{lemma:isnotzero}
  \begin{equation}
    (\delta_0-\delta_1) x \not=0 
    \quad \forall x\in e_1 (\C{}Q)-\{0\}
  \end{equation}
\end{lemma}
\begin{proof}
  To show that a given sum-of-paths is not zero we just have to show
  that the ``smallest'' path is nonzero, i.e. it suffices to show that
  $\delta_0 x \not=0$. But prefixing each path with $\delta_0$ acts
  injectively on the normal forms.
\end{proof}

\subsection*{The Tilting Complex}

Again I will use theorem~\ref{thm:tilting} to find a BDS-dual
quiver. I claim that $T\eqdef T_1\oplus T_2 \oplus T_3 \oplus T_4$ is
a tilting complex, where
\begin{equation}
  \label{eq:T1234defToric}
  \begin{split}
      T_1 \eqdef \quad
      \xymatrix@=25mm@!0{
        0 \ar[r] & \underline{P_4 \oplus P_4} 
        \ar[r]^-{
          \left(
          \begin{smallmatrix}
            \delta_0 \\ -\delta_1
          \end{smallmatrix}
          \right)} & 
        P_1 \ar[r] & 0 
      }
    \\
      T_2 \eqdef \quad
      \xymatrix@=25mm@!0{
        0 \ar[r] & \underline{P_2} \ar[r] & 0 \ar[r] & 0 
      }
    \\
      T_3 \eqdef \quad
      \xymatrix@=25mm@!0{
        0 \ar[r] & \underline{P_3} \ar[r] & 0 \ar[r] & 0 
      }
    \\
      T_4 \eqdef \quad
      \xymatrix@=25mm@!0{
        0 \ar[r] & \underline{P_4} \ar[r] & 0 \ar[r] & 0 
      }
  \end{split}
\end{equation}
Again $T1$ has homology at position $-1$ and $0$, so again this
tilting complex cannot simply be rephrased as a tilting module. Now
first I have to show that the $\Hom$'s between shifted $T_a$'s vanish.
The nontrivial cases are
\begin{descriptionlist}
\item[$\Hom(T_2\T{-1},T_1)=\Hom(T_3\T{-1},T_1)=\Hom(T_4\T{-1},T_1)=0$]
  because possible chain maps are homotopic to zero.
\item[$\Hom(T_1\T{1},T_2)=\Hom(T_1\T{1},T_3)=\Hom(T_1\T{1},T_4)=0$]
  because there are no chain maps, for example take
  \begin{equation}
    \vcenter{\xymatrix{
      \strut  
      0   \ar[r]      & 
      P_4\oplus P_4         
        \ar[r]^-{
          \left(\begin{smallmatrix}
            \delta_0 \\ -\delta_1
          \end{smallmatrix}\right)} 
        \ar[d] &
      \underline{P_1}
                 \ar[r] \ar[d]^{x} &
      0 
    \\
      \strut 
      0    \ar[r] & 
      0    \ar[r] & 
      \underline{P_2}  \ar[r] & 
      0
    }}
  \end{equation}      
  then by lemma~\ref{lemma:isnotzero} commutativity requires $x=0$.
\item[$\Hom(T_1\T{1},T_1)=0$] again since there is no chain map:
   \begin{equation}
    \vcenter{\xymatrix{
      \strut  
      0   \ar[r]      & 
      P_4\oplus P_4         
        \ar[r]^-{
          \left(\begin{smallmatrix}
            \delta_0 \\ -\delta_1
          \end{smallmatrix}\right)} 
        \ar[d] &
      \underline{P_1}
                 \ar[r] \ar[d]^{x} &
      0 \ar[r] \ar[d] & 
      0
    \\
      \strut 
      0   \ar[r]      & 
      0   \ar[r]      & 
      \underline{P_4\oplus P_4}
        \ar[r]^-{
          \left(\begin{smallmatrix}
            \delta_0 \\ -\delta_1
          \end{smallmatrix}\right)} &
      P_1 \ar[r] & 
      0 
    }}
  \end{equation}      
  While the right square can be made commutative with $x\not=0$, the
  left square can not (again by lemma~\ref{lemma:isnotzero}).
\end{descriptionlist}
Second I have to show that $T_1,\dots,T_4$ generate the whole
$\Db(\C{}Q\catmod)$. Again this is so because we can generate $P_1$ by
the mapping cone of $T_1\to T_4\oplus T_4$.

\subsection*{The Dual Quiver}

Determining the dual quiver is again technically more complicated
because there are infinitely many maps from each $T_a$ to each $T_b$.
So writing down infinitely many maps, almost all of which can be
eliminated by relations, is obviously a bad idea. Instead we must be
careful to identify the elements of $\End(T)$ which do not factor
through other endomorphisms.

First consider $T_2$, $T_3$ and $T_4$. Since they are just stalk
complexes the homomorphisms are generated by the paths in $Q$. They
are summarized in table~\ref{tab:HomToric}, where again I will use the
notation $[x_1,x_2]$ (as in table~\ref{tab:BDtoyHom}) to denote the
two nontrivial chain maps between the summands of $T$.
\begin{table}[htbp]
  \centering
  {   
  \begin{tabular}{c|cccc}
    $A\backslash B$ & $T_1$ & $T_2$ & $T_3$ & $T_4$ \\ \hline
    $T_1$ &
      $\left[ \left(\begin{smallmatrix} 1 & 0 \\ 0 & 1
              \end{smallmatrix}\right)
              ,1 \right]$ &
      & % none $\left[,\right]$ &
      & % none $\left[,\right]$ &
      \begin{math}\begin{array}{r@{\eqdef}l}
        y_{9}  & \left[ \left(\begin{smallmatrix} 0 \\ 1
                        \end{smallmatrix}\right)
                        ,0 \right] \\
        y_{10} & \left[ \left(\begin{smallmatrix} 1 \\ 0
                        \end{smallmatrix}\right)
                        ,0 \right]
      \end{array}\end{math} 
    \\ \\
    $T_2$ &
      \begin{math}\begin{array}{r@{\eqdef}l}
        y_{7} & \left[ \left(\begin{smallmatrix} 
                         \beta_1 \gamma_0 ~& \beta_0 \gamma_0
                       \end{smallmatrix}\right)
                       ,0 \right] \\
        y_{8} & \left[ \left(\begin{smallmatrix} 
                         \beta_1 \gamma_1 ~& \beta_0 \gamma_1
                       \end{smallmatrix}\right)
                       ,0 \right] 
      \end{array}\end{math} &
      $\left[1,0 \right]$ & 
      \begin{math}\begin{array}{r@{\eqdef}l}
        y_1 & \left[\beta_0,0 \right] \\
        y_2 & \left[\beta_1,0 \right] 
      \end{array}\end{math} &
      % none $\left[,\right]$ 
    \\ \\
    $T_3$ &
      & % none $\left[,\right]$ &
      & % none $\left[,\right]$ &
      $\left[1,0 \right]$ & 
      \begin{math}\begin{array}{r@{\eqdef}l}
        y_3 & \left[\gamma_0,0 \right] \\
        y_4 & \left[\gamma_1,0 \right] 
      \end{array}\end{math} 
    \\ \\
    $T_4$ &
      & % none $\left[,\right]$ &
      \begin{math}\begin{array}{r@{\eqdef}l}
        y_{5}  &  \left[\delta_0 \alpha_0,0 \right] \\
        y_{6}  &  \left[\delta_0 \alpha_1,0 \right] \\
        y_{11} &  \left[\delta_1 \alpha_0,0 \right] \\
        y_{12} &  \left[\delta_1 \alpha_1,0 \right] 
      \end{array}\end{math} &
      & % none $\left[,\right]$ 
      $\left[1,0 \right]$ 
  \end{tabular}
  } % end renewed baseline/array stretch
  \caption{Summary of generators of $\Hom(A,B)$ in $\Db(\C{}Q\catmod)$.}
  \label{tab:HomToric}
\end{table}
More difficult are the $\Hom$'s involving $T_1$. I discuss the cases
shown in table~\ref{tab:HomToric}:
\begin{descriptionlist}
\item[$\Hom(T_1,T_4)$:] They are obviously generated by the projection
  on the first or second $P_4$ summand.
\item[$\Hom(T_1,T_2)~\text{and}~\Hom(T_1,T_3)$:] Every such morphism
  factors through $\Hom(T_1,T_4)$, so I do not include them into the
  list of generators.
\item[$\Hom(T_4,T_1)$:] The simplest guess $[\left(\begin{smallmatrix}
      1 ~& 0 \end{smallmatrix}\right), 0]$ does not work since it is
  not a chain map. One has to ``go around the square'' once, so one
  can use a relation in the path algebra to generate a chain map, like
  \begin{equation}
    \begin{gathered}
    \vcenter{\xymatrix{
      \strut  
      0   \ar[r]      & 
      \underline{P_4}
        \ar[r] 
        \ar[d]_{
          \left(\begin{smallmatrix}
            \delta_0 \alpha_0 \beta_1 \gamma_0 ~& 
            \delta_0 \alpha_0 \beta_0 \gamma_0
          \end{smallmatrix}\right)} &
      0 \ar[r] \ar[d] &
      0 
    \\
      \strut 
      0    \ar[r] & 
      \underline{P_4\oplus P_4}
        \ar[r]^-{
          \left(\begin{smallmatrix}
            \delta_0 \\ -\delta_1
          \end{smallmatrix}\right)} &
      P_1  \ar[r] &
      0
    }}
    \\ % split
      \left(\begin{smallmatrix}
        \delta_0 \alpha_0 \beta_1 \gamma_0 ~& 
        \delta_0 \alpha_0 \beta_0 \gamma_0
      \end{smallmatrix}\right) 
      \left(\begin{smallmatrix}
        \delta_0 \\ -\delta_1
      \end{smallmatrix}\right) 
      = \delta_0 \alpha_0 (
        \beta_1 \gamma_0 \delta_0  - \beta_0 \gamma_0 \delta_1 )  
      = 0
    \end{gathered}
  \end{equation}
  However such a morphism factors through the $\Hom(T_4,T_2)$ to be
  discussed below, and is therefore not included in the
  table~\ref{tab:HomToric} of generators.
\item[$\Hom(T_2,T_1)$:] The possible chain maps 
  \begin{equation}
    \vcenter{\xymatrix{
      \strut  
      0   \ar[r]      & 
      \underline{P_2}
        \ar[r] 
        \ar[d]_{
          \left(\begin{smallmatrix}
            x_1 ~& x_2
          \end{smallmatrix}\right)} &
      0 \ar[r] \ar[d] &
      0 
    \\
      \strut 
      0    \ar[r] & 
      \underline{P_4\oplus P_4}
        \ar[r]^-{
          \left(\begin{smallmatrix}
            \delta_0 \\ -\delta_1
          \end{smallmatrix}\right)} &
      P_1  \ar[r] &
      0
    }}
  \end{equation}
  have to satisfy
  \begin{equation}
    \begin{pmatrix}
      x_1 & x_2
    \end{pmatrix}
    \begin{pmatrix}
      \delta_0 \\ -\delta_1
    \end{pmatrix}
    = 0
  \end{equation}
  The minimal (path length $2$) solutions are 
  \begin{equation}
    \begin{pmatrix}
      x_1 & x_2
    \end{pmatrix}    
    \in \Span_{\C{}}\Bigg( 
    \begin{pmatrix}
      \beta_1 \gamma_0 ~& \beta_0 \gamma_0
    \end{pmatrix}    
    ,~
    \begin{pmatrix}
      \beta_1 \gamma_1 ~& \beta_0 \gamma_1
    \end{pmatrix}    
    \Bigg)
  \end{equation}
  and all longer (path length $6$, $10$, \dots) solutions factor
  through those.
\item[$\Hom(T_1,T_1)$ and $\Hom(T_3,T_1)$:] These do not yield new
  generators for the same reason as $\Hom(T_4,T_1)$.
\end{descriptionlist}

\subsection*{The Relations in the Dual Quiver}

I fixed a nice set of generators in table~\ref{tab:HomToric}, but they
are not independent. Using the relations in $Q$ it is easy to check
the following $10$ relations in $\End(T)$:
\begin{equation}
  \label{eq:ToricRel1}
  \begin{array}{r@{=}l@{\hspace{5mm}}
                r@{=}l@{\hspace{5mm}}
                r@{=}l@{\hspace{5mm}}
                r@{=}l@{\hspace{5mm}}
                r@{=}l@{\hspace{5mm}}}
    y_{2} y_{4}  & y_{8} y_{10} &
    y_{1} y_{4}  & y_{8} y_{9}  &
    y_{4} y_{11} & y_{3} y_{12} &
    y_{5} y_{8}  & y_{6} y_{7}  &
    y_{6} y_{2}  & y_{12} y_{1} 
  \\
    y_{2} y_{3}  & y_{7} y_{10} &
    y_{1} y_{3}  & y_{7} y_{9}  &
    y_{11} y_{8} & y_{12} y_{7} &
    y_{4} y_{5}  & y_{3} y_{6}  &
    y_{5} y_{2}  & y_{11} y_{1}
  \end{array}
\end{equation}
The above relations hold already on the level of chain maps. But
in the derived category we also have to identify homotopic maps,
leading to additional relations in $\End(T)$. The two additional
relations are 
\begin{equation}
  \label{eq:ToricRel2}
  y_{10} y_{6} = y_{9} y_{12}
  \qquad
  y_{10} y_{5} = y_{9} y_{11}  
\end{equation}
For example to show that $y_{10} y_{6} = y_{9} y_{12}$ one has to show
that the difference of the chain maps is homotopic to zero. The
homotopy is straightforward to find:
\begin{equation}
  y_{10} y_{6} - y_{9} y_{12} \sim 0: \quad
  \vcenter{\xymatrix{
    \strut  
    0   \ar[r]      & 
    \underline{P_4\oplus P_4}
      \ar[r]^-{
        \left(\begin{smallmatrix}
          \delta_0 \\ -\delta_1
        \end{smallmatrix}\right)} 
      \ar[d]_-{
        \left(\begin{smallmatrix}
          \delta_0 \alpha_1 \\ -\delta_1 \alpha_1
        \end{smallmatrix}\right)}  &
      \ar@{.>}[dl]^{\alpha_1}
    P_1
      \ar[r] \ar[d] &
    0 
  \\
    \strut 
    0    \ar[r] & 
    \underline{P_2}  \ar[r] & 
    0 \ar[r] & 
    0
  }}
\end{equation}      
Together the relations eq.~\eqref{eq:ToricRel1}
and~\eqref{eq:ToricRel2} are precisely the F-terms for the
superpotential in the toric dual quiver in eq.~\eqref{eq:Qtoricdual}.
So in this case strong toric duality is a BDS-duality, and the
equivalence of derived categories is induced by the tilting complex
$T$ above.

\section{Non-Dualities and Invariants of the Derived Category}
\label{sect:invariants}

Clearly it is desirable to have simple criteria to check whether the
derived categories can at all be derived equivalent.

\subsection{Global Dimension}
\label{sect:gldim}

As the simplest example of a conserved quantity I would like to
consider the (finiteness of the) global dimension of the quiver. Take
all possible modules of the path algebra, then for each one there is a
minimal projective resolution. The maximal length is the global
dimension of the quiver path algebra.

This is useful because of the following (see~\cite{HappelBook}):
\begin{theorem}
  Let $\C{}Q_1$ and~$\C{}Q_2$ be finite dimensional algebras. If
  $\Db(\C{}Q_1\catmod) \simeq \Db(\C{}Q_2\catmod)$ and $\C{}Q_1$ has
  finite global dimension then $\C{}Q_2$ also has finite global
  dimension.
\end{theorem}
With the help of this theorem we can immediately show that ``closing
up the loop'' in $Q_2$ of eq.~\eqref{eq:BDtoyQ} does change the
derived category: Let 
\begin{equation}
  Q_2 \eqdef \quad
  \begin{displayarray}{c}
    \xymatrix{ 
      \qnode{1} & 
      \qnode{2} \ar[l]_{\phi_{21}} & 
      \qnode{3} \ar[l]_{\phi_{32}}
    }
    \\
    \phi_{32}\phi_{21}  =0
  \end{displayarray}
  \qquad\qquad
  Q_3 \eqdef
  \vcenter{\xymatrix@R=0.8cm@C=0.3464cm{ & 
    \qnode{3} \ar[ld]_{\phi_{31}} \\
    \qnode{1} \ar[rr]^{\phi_{12}} & & 
    \qnode{2} \ar[lu]_{\phi_{23}} 
  }}
  \begin{array}{c}
    \phi_{12} \phi_{23} =0 \\ 
    \phi_{23} \phi_{31} =0 \\ 
    \phi_{31} \phi_{12} =0 
  \end{array}
\end{equation}
and note that $\gldim \C{}Q_2=2$, $\gldim \C{}Q_3=\infty$. Therefore
$\Db(\C{}Q_2\catmod)\not\simeq\Db(\C{}Q_3\catmod)$.

Note that it is well-known that the global dimension itself (if
finite) is not preserved under equivalence of derived categories, only
whether it is finite or infinite. It would be nice if one had a
physical argument why this should be invariant.

% FIXME: what about calabi-yau, dim=3?

\subsection{\Ktheory}
\label{sec:ktheory}

Let us for the moment go back to the string theoretic motivation and
think of the $\Ncal=1$ SYM as the low energy effective action of
\Dbranes{} on a \CYm{}. Now consider two possible brane setups: In one
case you allow only for (arbitrary numbers of) branes wrapping a
single fixed cycle, and in the other case you allow branes wrapping
all possible cycles. Will the two $\Ncal=1$ theories be dual? Of
course they should not, the second case should contain a lot more
physical information.

Put differently, in the first case I considered only multiples of one
particular \Dbrane{} charge, while in the second I allowed all
possible charges. Two \Dbrane{} categories should only be dual if
their \Ktheory{} lattice is the same.

This fits very nicely with the mathematics of derived categories
(see~\cite{HappelBook}). To each derived category $\Db(A)$ we can
associate the Grothendieck group $K_0(A)$ by modding out
\begin{enumerate}
\item Isomorphism
\item Elements of the form $[X]-[Y]+[Z]$ for each triangle $X\to Y\to
  Z\to X\T{1}$.
\end{enumerate}
and this is the same as the \Kgroup{} of $A$ itself. So the derived
category contains all the information of the \Kgroups.

Now again we can forget about the \Dbranes{} and work only with the
derived quiver representations. The \Kgroups{} of $\Db(\C{}Q\catmod)$
is the same as the \Ktheory{} of the path algebra. The \Ktheory{} of
the algebra is generated by the projective modules, which have in our
case a quite simple structure: They are in one-to-one correspondence
with the nodes of the quiver, theorem~\ref{thm:projective}. Thus
\begin{equation}
  K_0( \C{}Q ) = \Z{}^{ \# \text{of nodes of~}Q}
\end{equation}
So we see immediately that the weakly toric dual quivers of
eq.~\eqref{eq:Qtoricdual} cannot have equivalent derived categories
since the number of nodes is different.

Also note that $K_0(\C{}Q)$ is always free, i.e. there is no torsion
($\Z{n}$) subgroup. One should not expect that one can always
understand the derived category of coherent sheaves by simply studying
quivers.

\section*{Conclusions}
\addcontentsline{toc}{section}{Conclusions}

I have reviewed the conjecture of Berenstein and
Douglas~\cite{BerensteinDouglas} which can be seen as a precise and
unambiguous definition of Seiberg duality.  Despite the fearsome
mathematical language involved it is impressive that it can be stated
in one single line.

To actually apply the definition one has to unravel the words, which
can lead do quite intricate computations. It is already nontrivial to
check that it gives a reasonable answer for the original toy quiver
of~\cite{BerensteinDouglas}.

But to actually check the proposal one needs examples for dual gauge
theories, that is quivers with superpotential. I gave a fairly simple
example of such a BDS-dual quiver pair, and explicitly checked the
equivalence of derived categories. For this we had to work really in
the derived category and go beyond tilting modules.

Although it sounds scary it is actually quite feasible to show the
equivalences of derived categories. To demonstrate the power of this
approach I then tackled a technically much more complicated problem: I
showed explicitly that a pair of toric dual quivers is BDS-dual. While
the equivalence of derived categories was expected on general grounds
it is a very nice check that the algorithm to associate the quiver to
the toric variety is correct.

Finally I mentioned some invariants under derived equivalence, which
are very useful if one wants to show that two quivers are \emph{not}
BDS-dual.

\section*{Acknowledgments}

I would like to thank the organizers and lecturers of the ``Clay
Mathematics Institute/ Isaak Newton Institute School on Geometry an
String Theory'' and especially the derived discussion group for
sparking my interest in derived categories.

\bibliographystyle{JHEP}
\renewcommand{\refname}{Bibliography}
\addcontentsline{toc}{section}{Bibliography}
\bibliography{seiberg}

\end{document}